# A computational account of dreaming: learning and memory consolidation

**Qi Zhang**

Sensor System Madison, WI, USA
qzhang_sensor@yahoo.com

**Abstract**

A number of studies have concluded that dreaming is mostly caused by randomly arriving internal signals because “dream contents are random impulses”, and argued that dream sleep is unlikely to play an important part in our intellectual capacity. On the contrary, numerous functional studies have revealed that dream sleep does play an important role in our learning and other intellectual functions. Specifically, recent studies have suggested the importance of dream sleep in memory consolidation, following the findings of neural replaying of recent waking patterns in the hippocampus. The randomness has been the hurdle that divides dream theories into either functional or functionless. This study presents a cognitive and computational model of dream process. This model is simulated to perform the functions of learning and memory consolidation, which are two most popular dream functions that have been proposed. The simulations demonstrate that random signals may result in learning and memory consolidation. Thus, dreaming is proposed as a continuation of brain’s waking activities that processes signals activated spontaneously and randomly from the hippocampus. The characteristics of the model are discussed and found in agreement with many characteristics concluded from various empirical studies.

## A computational account of dreaming: learning and memory consolidation

## Introduction

Dreaming refers to the subjective conscious experiences we have during sleep. The experience is vivid, intense, bizarre, and is hard to recall. Various studies have concluded that dream sleep may help us in learning (e.g., Greenberg & Pealman, 1974; LaBerge, 1985; Hennevin, et al., 1995; Smith, 1995), and may be a perceptible embodiment of a dreamer's conceptions (Hall, 1953). Findings of the correlation between REM (rapid eye movement) sleep and waking learning have suggested that dream sleep may play an important role in learning and memory consolidation (e.g., Fishbein, 1970; Pearlman, 1971; Bloch, et al., 1979; Winson, 1985). Similar findings have been also concluded in more recent psychophysiological studies, although only these initiating studies are cited.

Declarative memory of long-term memory is generally divided into episodic memory and semantic memory. The formal is the memory of past experiences, and the latter is the memory of conceptual knowledge (Tulving, 1972). Memory consolidation is a neural process by which episodic memory becomes independent of the hippocampal complex and is consolidated into the neocortex (Squire & Alvarez, 1995). The complex, including the hippocampus and its surrounding areas, is considered a critical region in retaining recent episodic memory or its traces. On the contrary, the general neocortex is considered the place where semantic memory is stored. Findings from neural recording, which reveal the replaying of recent waking patterns of neuronal activity within the hippocampus during sleep, have further reinforced the view that dreaming may play an important part in memory consolidation. This reactivation of hippocampal cells has been recorded in rats (e.g., Pavlides & Winson, 1989; Wilson & McNaughton, 1994) and in humans (Staba, et al., 2002); in SWS (slow wave sleep) that dominates Non-REM sleep (Pavlides & Winson, 1989; Wilson & McNaughton, 1994) and in REM sleep (Poe, et al., 2000; Louie & Wilson, 2001). Furthermore, synchronized activity in the hippocampus and neocortex in sleep is also reported and attributed to memory consolidation into the neocortex (Battaglia1, et al., 2004).

In some studies (e.g., Foss,e et al., 2003; Pavlides & Winson, 1989; Schwartz, 2003), dream sleep has been directly associated with the hippocampal firings for the possible link between the cognitive activity of brain and the activation of stored episodic memory. This association is reasonable when we consider the fact that both vivid dreams and thought-like experiences can be recalled in >70% of REM awakenings (Hobson, 1988), and >48% of Non-REM awakenings (Nielsen, 2000), respectively. However, this association has an obstacle. Carefully looking into dream contents, it is often concluded that dreams are more or less random thoughts (Hobson & McCarley, 1977; Foulkes, 1985; Wolf, 1994). A recent study by Fosse, et al. (2003), which is focused on the correlation between daily experiences and dream contents, again confirms the randomness nature of dreams. This study found that daily experiences are replayed in the form of segments, rather than entire episodes, during REM sleep. In other words, daily experience is replayed more in random fashion and less in sequential fashion in dreams. The randomness has led to the proposal of the activation-synthesis model (Hobson & McCarley, 1977; Hobson, 1988; Hobson, et al., 2000). The model states: dreams are caused by random signals arising from the pontine brainstem during REM sleep; the forebrain then synthesizes the dream and tries its best to make sense (i.e., dream images) out of the nonsense (i.e., random impulses) it is presented with. In short, the dream randomness has been used against some proposed intellectual functions of dreaming, and has divided dream theories into functional and functionless. Therefore, whether or not dreaming has intellectual functions depends on whether "random impulses" can lead to intellectual consequences, e.g., learning and memory consolidation.

There have been few reported computational studies of dream simulations by artificial neural networks. One simulation suggests that dreaming is the "reverse-learning" process to remove so-called spurious memories (i.e., useless and old memories) in order to avoid overload of the brain (Crick & Mitchison, 1983). A similar simulation, however, indicates that the useless and old memory is actually increased after the "reverse-learning" of the simulated dreaming process, and suggests that we dream to roughen up our "memory space" (Christos, 2003). In either case, it is said (Botman & Crovitz, 1989; Domhoff, 1996) that the conclusions are generally disassociated with what has been found about dreaming, and are typically interpreted to contradict the psychoanalytic account of dreams.

A cognitive and computational model of dreaming is presented. This model is developed from a previous construct (Zhang, 2005) of a learning system. In this present study, dreams of the computational system are performed. The outcomes of dreaming are examined in terms of "naming" and "picture drawing," which are typical tasks in the tests for semantic memory. The characteristics of the model are discussed and found in agreement with many empirical findings from dream studies.

## The construct of an artificial intelligence dreamer

### *A brief revisit of the learning system*

Knowledge can be learned from experience. How learning occurs and how new knowledge is associated with prior knowledge, are questions yet to be answered. A previous study (Zhang, 2005) presents a cognitive learning system, namely AI counter, which can learn to count. That system is constructed based on two rules: (1) a concept (represented by a common feature) is learned when the common feature is abstracted and generalized; and (2) new learning has to rely on prior learning if the newer knowledge is an extension of the prior knowledge. The learning system is built with a multi-level structure of information processing, which is in fact the system shown in Figure 1, except for the "hippocampal memory". In the system, the base level of cognition is called "single memory" that stores and reacts to one piece of an entire external input. Three single memories are then organized into a "memory triangle" (shown in Figure 2) that stores, generalizes and reacts to one common feature carried by an external input. At a higher level, a subsystem is formed by several interactive and interlocked memory triangles. There are five kinds of signals throughout the entire system from the level of a single memory to the level of a subsystem, as shown Figure 1 and 2. Among them, three (excitation, coordination and interlock signals) are for internal communication, and two (symbol or representation signals) are for external communication (see also the signal keys in Figure 1 and 2). Inspired by the findings of the "split brain" (Myers & Sperry, 1953; Sperry 1982)—our left hemisphere favors symbolic process, and our right hemisphere favors perceptual process, two such subsystems are paired up to form the entire learning system. One subsystem is dedicated to symbols and the other is to the symbols' representations (i.e., the perceptual meanings). The two subsystems communicate with each other via a bundle of inter-subsystem signals that mimics the corpus callosum in connecting the two hemispheres of a brain (Myers & Sperry, 1953).

| Figure 1 | | Figure 2 |
|---|---|---|

Concept is an abstract idea or a mental symbol. Words are not concepts but signs for concepts. According to Immanuel Kant (1800), a concept is a common feature or characteristic, and concepts are abstracts in that they omit the differences of the things in their extension, treating them as if they were identical. The AI counter is so constructed that it can abstract common features from given examples, and can link the common features to their given symbols after learnings have occurred. In order for the AI counter to be able to tally, it has to learn three concepts that are associated with three common features. The three features are about the meaning of "zero", "one" and "tally", and are carried by those eight pairs of examples in either Table 1 or Table 2. For example, the 4th example, in Table 1, is an input pair of a symbol "I", and a single-peaked representation. This example "tells" the learning system that the symbol "I" is the sign for a one-peaked representation. Further, the 6th example is an input pair of the symbol "I", and another single peaked representation. However, this single peak looks different from that of the 4th representation input. Thus, this example "tells" the system that the symbol "I" is also the sign for a different one-peaked representation. The learning system has to "omit the differences of the things in their extension" and to abstract what is in common from given examples. In here, the differences are the widthes of different single-peaks, while the common feature is “one” peak.

The three concepts are carried by the eight examples in either Table 1 or Table 2. For example, the input pairs of example 1 through 3 are about the concept of “zero”, whose representation input is a flat line with no peak in it; the pairs of example 4 through 6 are about the concpet of “one” , and whose presentation input has a single peak. After the system has learned these three concepts, it can recognize external stimuli and count them, even if the stimuli have never been experienced before, as long as they are consistent with the concepts that have been learned. When receiving a symbol input, the system answers it with a representation of the symbol; when receiving a representation input, it responds with the symbol of the representation. The former operation is equivalent to “object drawing” and the latter to “object naming” that are terms of typical tasks for semantic memory.

However, the learning rate of the system is dependant of training examples. It only needs one round of training with the examples given in Table 1, which has been demonstrated in the previous study, but it has to be trained for six rounds with the examples given in Table 2. Both Table 1 and 2 contain the same eight examples. The only difference is that the examples in Table 1 are listed in sequence for the meanings of “zero”, “one” and “tally”, and are listed from simple to complex. Comparably, the examples in Table 2 are not listed in the sequence from simple to complex. Based on the proposed rule of learning (Zhang, 2005), new learning has to rely on prior learning if the newer knowledge is an extension of the prior knowledge. Thus, the learning system cannot learn extended knowledge if its base knowledge has not already been learned, which is the reason why it learns slower with the examples in Table 2. When it cannot learn from an example, it simply lets the external experience pass through like nothing has happened. Therefore, it has to experience those examples many more times before learning is finally realized. What if the learning system had a subsystem that could store external experience as its "episodic memory" for later reprocessing? Comparably, humans, and most mammals, do have the capability to remember rapidly what has happened and to retain this as episodic memory after a single experience.

| Table 1 | Table 2 |
|---|---|

*An AI dreamer*

In humans and almost all mammals, there is such a memory system, namely the hippocampal complex, which has been revealed to have the function of immediate storage (e.g., Tulving, 1972; Squire, 1992; McClelland, et al., 1995). When the complex is damaged, the capacity of retaining episodic memory is impaired (Scoville & Milner, 1957) and patients become amnesic (e.g., Squire & Alvarez, 1995). Deterioration of this complex has been commonly found in patients with Alzheimer's disease for whom poor retention of episodic memory is the first sign.

Based on the function of the biological hippocampal complex, a "hippocampal memory" subsystem is added to the learning system of AI counter. As shown in Figure 1, the subsystem has a number of "memory cells", and each cell is designated to store a segment of an external experience. All of the cells are interlocked so that the sequenced segments of an external experience are stored in their original order of occurrence. The interlocking mechanism has similar functions as the hippocampus in terms of sequential learning (e.g., Levy, 1996) when the variable is time, and spatial navigation (e.g., Muller, et al., 1987) when the variable is distance. Thereby, all stored segments of an external experience can be re-accessed and replayed in their originally occurring sequences (to be demonstrated). The segments can also be fired randomly, which becomes the source of random signals in dream simulations of this study.

The system, shown in Figure 1, is named AI dreamer, which is made up of two cognitive sections. One section is the AI counter of the previous study, which is also called the "learning center" in later discussions for convenience. And, the other is the "hippocampal memory" that stores its experience and allows the experience to be re-accessed. In here, the "conceptual knowledge" and "stored experience" are compared with the concepts of semantic and episodic memories defined by Tulving (1972), respectively. This two-sectioned construct also agrees with an understanding that the brain has a complementary memory system of knowledge and experience, each of which uses different computational strategies for storing information (e.g., O'Keefe & Nadel, 1978; McClelland, et al., 1995).

Besides the added hippocampal memory, a "mode selector" is also added that can set the system either to waking mode in order to receive external stimuli, or set the system to dreaming mode to process internal signals that are fired randomly from the hippocampal memory. Similar as for the learning system of the AI counter, the cognitive system (namely AI dreamer) shown in Figure 1 is also built in the LabView language.

**Simulations of the AI dreamer**

The external experience for the system to learn is a series of input pairs given in Table 2. Each pair of the input consists of a symbol and its representation. When the system learns, it takes in an input pair via separated channels, $S_{in}$ and $R_{in}$, shown in Figure 1. The representation input carries the meaning of the symbol input, while the symbol is a "name" that symbolizes the representation. The term "meaning" in here means that the input carries the conceptual knowledge that can be perceived and understood in a real world. We can see from the Table, that a representation input may be a flat line or may come with a number of peaks. Every peak represents one "object", while a flat line means "no" object. When we see an input has three peaks, the meaning of three peaks is to be perceived by the learning center. Those input pairs in either Table 1 or 2 carry the same knowledge (i.e., "zero", "one" and "tally") necessary for the system to learn how to tally. The only difference between the two tables is in the given sequence of the input pairs.

Firstly, in training phase, the system is set to waking mode to receive all given pairs in Table 2 in their given sequence. This system only receives those pairs once, but it must "remember" the entire experience thereafter. This is comparable with the fact that human can remember what has happened after a single experience, and this property of rapid remembering is demonstrated through the "serial recall" (that is also a standard test in memory study) shown in Table 3.

When an input pair of symbol and representation arrives during the training, the symbol is forwarded to the symbol subsystem, while the representation is forwarded to the representation subsystem. In turn, the inputs are delivered to the memory triangles and single memory at lower cognitive levels. The learning center may learn from, react to, or reject the input pair. What the learning center eventually does is a result of interactions amongst all its enclosed single memories, each of which it also has the potential of learning from, reacting to or rejecting. If the center cannot learn from an input pair, it rejects and forwards the pair to the hippocampal memory via the two pathways as indicated in Figure 1. The pair is then stored in one of the memory cells as a part of "episodic memory".

After the training phase, the learning center is only able to learn from the 6$^{th}$ pair, and it has to store the rest of the pairs into the hippocampal memory. This is because the 6$^{th}$ pair is the starting point of the most basic knowledge of tally—a flat baseline that means there is "no" object. Learning from the 6$^{th}$ pair is only the beginning to learn the baseline. Thereafter, the learning center has to learn from the 5$^{th}$ pair and then the 4$^{th}$ pair before it can fully generalize the meaning of a baseline (i.e., the meaning of "zero"). Only when it knows what is a baseline, it is able to subtract one "object" (i.e., one peak) from the baselines of the 3$^{rd}$, 8$^{th}$, and 2$^{nd}$ pairs, and to learn the property of the subtracted "one object". After it has learned the meaning of "one", it then can subtract more peaks and abstract the meaning of tally in subsequent learnings of the 7$^{th}$ and 1$^{st}$ pairs. This progressive learning is the result of the mechanism proposed in the previous study that new learning has to rely on prior learning when the newer knowledge is an extension of the prior knowledge. All these subsequent learning steps, however, does not occur in the training phase, and only occur in the dream phase in which the stored examples are randomly fired and processed by the learning center.

After the initial external experience, the system is set to dream mode to allow the learning center to reprocess those stored experiences fired from the hippocampal memory. During dreaming, the learning center acts in the same way as it is “awake”: it learns from an input pair if the input’s prior knowledge has already been learned; it responds to an input if the meaning of the input has already been learned; and it rejects an input pair if the pair’s prior knowledge has not been learned.

During dreaming, the stored segments of experience are randomly fired, which is intended to simulate the random nature of dreaming that has been concluded by a number of dream studies. The randomness has strongly challenged the view that dreaming has any intellectual function. For example, dreaming is thought only a neural process by which our brain produces dream imagery from noisy signals (Hobson, 1988; Hobson & McCarley, 1977). The central point of the challenge is based on the assumption that randomly assembled information hardly contributes to cognition and knowledge. But, the assumption may not be true. From a philosopher’s viewpoint, knowledge can be learned through experience of a direct causal (experience-based) interaction between a person and the object that the person is perceiving (Russell, 1913). On the other hand, it has also been argued that true knowledge (or at least the most important knowledge) is essentially independent of sensory experience (Locke, 1690). If both statements are true, it may be said that knowledge can be acquired through experience, but the acquisition process is not necessarily dependent on an exact experience. Random firing may be suficient enough to provide an alternative to the exact replay of an actual experience in knowledge aquisition.

Table 3

Table 3 lists some recorded streams of representation flows that are intercepted along the pathway from the mode selector to the representation subsystem. The first simulation is a serial recall recorded in waking mode. If we simply attach all representation signals, given in Table 2 except the 6th input, into a "long representation signal", we will see that the "long representation signal" is same as the first recorded stream in Table 3. The rest streams are recorded in six different dream simulations of randomly activated "experiences". Similarly, each stream is made up of those randomly fired representation signals in the given dream of the AI dreamer. The results of learning, after the learning center has experienced the recorded dream streams, have also been indicated in the same table. As noted before, the input pairs in Table 2 carry all conceptual knowledge (including "zero", "one" and "tally") that are sufficient for the system to perform tallying after full learning. These concepts may be partially or fully learned in dreams under a given number of total firings. The efficiency of learning in a specific dream varies depending on how soon newer knowledge arrives after its required prior knowledge has been learned. The learning center may learn all of the conceptual knowledge in one dream (e.g., the first and second dreams in Table 3) or in many dreams (e.g., the series of four dreams in Table 3).

## Discussions

*Phenomenal and neurobiological properties of dreaming*

As we go to sleep, we slowly sink down into deeper stages of sleep (i.e., Non-REM or NREM sleep). After an hour or two, the first REM period begins and lasts a few minutes. Then, we sink back into NREM sleep again. This cycle occurs about every 90 minutes. Towards the end of the night, the REM periods get longer. NREM sleep alternates with REM sleep and includes all sleep apart from REM. Each NREM sleep has four stages corresponding to increasing depth of sleep as indicated by progressive dominance of the electroencephalography (EEG) by high-voltage, low frequency wave activity (Rechtschaffen & Kales, 1968). This low frequency wave is also called "slow wave", which dominates the deepest stages of NREM. Almost all mammals have the NREM–REM cyclic alternation in sleep, which suggests not only a shared mechanism across species but a universal functional significance (Hobson, et al., 2000).

Dreams can occur in both REM and NREM sleep. In 70-95% of awakenings from the REM state, normal subjects report that they have been dreaming whereas only 5-10% of NREM awakenings produce equivalent reports (Hobson, 1988). If there is no clear dream reported, 43%-50% of NREM awakenings can elicit reports of thought-like or mentation recall (Foulkes, 1962; Nielsen, 2000). If the mentation recall can also be accounted as the result of dreaming, it means the mind is mostly in dream states during sleep. There are more subtle differences between REM and NREM dreams. Reports from REM sleep awakenings are relatively longer, more perceptually vivid, more motorically animated, more emotionally charged, and less related to waking life than NREM reports (e.g., Foulkes, 1962; Antrobus, et al., 1987). In contrast to REM reports,

NREM reports are more thought-like and contain representations of current concerns more often than do REM sleep reports (Foulkes, 1962).

It has been noted earlier in the simulation of dreams that, in dream mode, the learning center is as active as it is in awaking mode. A similar characteristic has been concluded from neurobiological studies. It was stated (Hobson, et al., 2000) that, in REM sleep, the brain is almost as active as when awake, except for the sensory input and motor output, which are blocked, and for certain areas of the dorsolateral prefrontal cortex and the primary visual centers which are selectively deactivated. A number of functional imaging studies in humans have revealed a fairly consistent pattern of activities in REM sleep that appear to reflect dream processes (Maquet, et al., 1996; Braun, et al., 1997; Nofzinger, et al., 1997). Some important findings from the studies are: in REM sleep, the brainstem reticular formation is highly active; primary sensory areas (e.g., striate cortex for the visual system) are inactive; by contrast, extrastriate (visual) regions (as well as other sensory association sites) are very active; limbic and paralimbic regions (including the hippocampus, amygdala and anterior cingulate) are intensely activated; and widespread regions of the frontal cortex including the lateral orbital and dorsolateral prefrontal cortices show marked reductions in activity. These studies also show that the following areas are relatively less active in NREM than in REM sleep; they are the brainstem, midbrain, anterior hypothalamus, hippocampus, caudate, and medial prefrontal, caudal orbital, anterior cingulate, parahippocampal and inferior temporal cortices (Braun, et al., 1997).

Besides the fact that the general neocortex in dream sleep is almost as active as when awake, the highly activated limbic system, especially the hippocampus, is of special interest. The activated hippocampus coincides with the reactivation of recent waking patterns of neuronal activity within the hippocampus. The reactivation has been revealed in human and other mammals, and in REM and NREM sleeps, while the patterns of neuronal activity are generally thought to be associated with daily experience or episodic memory (Pavlides & Winson, 1989; Wilson & McNaughton, 1994; Poe, et al., 2000; Louie & Wilson, 2001; Staba, et al., 2002).

What could be the common functioning areas in both REM and NREM sleep? Functional imaging studies are able to indicate the activity intensities of various brain areas, but are not adequate to rule out the involvements of the less active areas. For example, bursts of firings have been recorded from neurons in the hippocampus during NREM sleep (Wilson & McNaughton, 1994), although brain imaging has indicated that this area is less active in NREM sleep. On the other hand, the findings that patients with large pontine lesions can still dream suggest that the brainstem is not the critical component for all dreams to occur, but rather it may only be important for REM dreaming (Solms, 1997). This study of pontine lesions is consistent with the finding from functional brain imaging that the pontine brainstem and anterior cingulate cortex is deactivated during NREM sleep (Braun, et al., 1997).

If only considering those activated areas in both REM and NREM sleep, it may be concluded that the commonly activated areas necessary for dreams to occur, at least, include: most of the general neocortex and of the limbic system (especially the hippocampus). This simplified activation pattern is compatible with the cognitive structure of the computational dreamer, which has a learning center (mimicking the conceptual learning function of the general neocortex) and a hippocampal memory (mimicking the storage and reactivation functions of episodic memory in the hippocampus). However, it must be pointed out that this system is only a simplified approach to explain some common properties of dreams in both REM and NREM sleep. By no means, is it thought to be able to enclose all dream properties of a brain.

*Psychophysiological characteristics of dreaming and dream mechanism*

The characteristics of dreaming that are concluded from many empirical studies can be employed to examine the proposed theory of dreaming and its simulations. The first is the characteristic of the randomness. This characteristic is concluded from or supported by empirical findings conducted by, e.g., Hobson and McCarley (1977), McCarley and Hoffman (1981), Hobson (1988), Reinsel et al. (1992), Williams et al. (1992), Revonsuo and Salmivalli (1995). These findings have been summarized by Hobson et al. (2000) as followings: (1) dream imagery can change rapidly, and is often bizarre in nature; (2) relative to waking and, when present, dreams often involve weak, post-hoc, and logically flawed explanations of improbable or impossible events and plots; and (3) dreams lack orientational stability; persons, times, and places are fused, plastic, incongruous and discontinuous.

The second one is the repetitive characteristic of dreams—repeated themes and repeated dreams (including recurrent dreams) are common among dreams (Domhoff, 1993). The proposed mechanism of conceptual learning states that new knowledge cannot be learned if its prior knowledge is not already learned. As a result, a segment of stored experience may by chance have to be fired multiple times before the learning center is able to learn from it. This multi-appearance of the same segment of stored experience can be commonly observed in those dream-source streams shown in Table 3. When the learning center is not able to fully abstract and generalize the knowledge from the random signals in one dream, more dreams become necessary before dream learning is thoroughly realized. That is the scenario of the multi-dream learning, shown also in Table 3, in which repeated "themes" and similar "dreams" are expected.

The lack of self-reflection and self-control are also the typical characteristics of dreaming. For example, self-reflection in dreams is generally found to be absent (Rechtschaffen, 1978) or greatly reduced (Bradley, et al., 1992); the dreamer rarely considers the possibility of actually controlling the flow of dream events (Purcell et al. 1986); volitional control is greatly attenuated in dreams (Hartmann, 1966). Similarly, this computational model of AI dreamer does not have a necessity of a central controller in regulating its learning process.

*The dream function of memory consolidation and learning*

The dream function of memory consolidation has long been suggested in various studies. Newer and relatively direct evidence comes from the neuronal activity of seemingly replay of waking-experience in the hippocampus in both REM and NREM sleep. Many researchers agree that such reactivation is experience dependent, and conclude that the reactivation is associated with the consolidation of episodic memory stored in the hippocampus (Pavlides & Winson, 1989; Wilson & McNaughton, 1994; Skaggs & McNaughton, 1996; Poe, et al., 2000; Staba, et al., 2002). Others who studied amnesia and hippocampal lesion have also expected the consolidation function in dream sleep. For example, Squire and Alvarez (1995) have argued that if episodic memory is revived constantly, dreaming may be the best answer to explain why the memory-consolidation process does not regularly intrude into our consciousness.

Neuronal recordings have suggested that daily experiences stored in the hippocampal complex can be reactivated or replayed during REM and NREM sleeps, but there are no indications about whether the replaying is sequential or random. Study of dream contents and their connections with daily experience may provide this detail. A recent study (Fosse et al. 2003) concluded that only 1.4% of dream reports contained the reproductions of experience ensembles; while 65% of dream elements are linked to fragments and features of waking events. This apparent contradiction between dreaming's lack of a fully episodic structure and its proposed role of the consolidation, has led to the suggestion (Fosse, et al., 2003; Stickgold, et al., 2001) that sleep only has its role in the consolidations of associated semantic memory and procedural memory, not episodic memory. In interpreting the data, however, some researchers have realized and suggested that the key is to understand why memories appear in dreams as fragments or partial episodes but also occasionally as complete replays (Schwartz, 2003; Nielsen & Stenstrom, 2005). The presented model offers one of the possible understandings: when random firing is the only fashion of the reactivation, daily experiences should only appear in the form of segments rather than the sequential replay of entire experience. Thus, experience related replay (not episode replay) should be in 65% of dreams, instead of 1.4%, if the study of dream contents of daily experience reproduction is reinterpreted.

Table 4

Memory consolidation is defined as the process by which memory becomes independent of the hippocampal region (Squire & Alvarez, 1995). The presented computational system can be employed to demonstrate the memory consolidation function through naming tests. This test is one of the standard memory tests in which the tested subject tells the name of a previously encountered object (or picture of the object) that is shown again. The hippocampal memory is removed after successful dream consolidation and is absent in the naming test, in order to demonstrate that the memory has become independent of the "hippocampal memory". The first three rows in Table 4 are summaries of three naming tests. In these tests, different "objects" (or pictures) are shown and the system answers with the correct names to each of the given objects. When compared with Table 2, one can see that the objects tested and the names answered are part of the inputs given in the table that have been experienced by the system previously. Since the tests are done after the "hippocampal memory" was removed, it indicates that, after dreaming, the "episodic memory" has been consolidated into the learning center and become free of the hippocampal memory.

It may be noticed, from those examples in Table 4, that the memory consolidation in the computational model is not a simple relocation of stored experiences, rather a learning process that gradually incorporates facts and events into an already existing framework of knowledge (e.g., McClelland, et al., 1995). The learning by consolidation is the natural outcome of the computational system that coincides with the understanding that mechanisms of the hippocampus and neocortex involved in the consolidation process, are different (e.g., O'Keefe & Nadel, 1978) however complementary (e.g., McClelland, et al., 1995). In other words, learning and memory consolidation can be seen as different sides of the same coin.

The function of dream learning is the most popular dream function that has been observed and concluded. The impact of dream learning is broad and can be often observed after dreams. For example, unprepared learning, more slowly mastered and difficult, is especially dependent on the quality of REM sleep (Greenberg & Pealman, 1974); learning tasks requiring significant concentration or acquisition of unfamiliar skills are followed by increased REM sleep (LaBerge, 1985); material learned during the day and consolidated over a night of sleep is recalled better the next morning (Hennevin, et al., 1995). The dream learning function can also be demonstrated with the presented system through both naming tests and "picture drawing" tests (also one of the standard memory tests) after the removal of the hippocampal memory. The last five rows in Table 4 are summaries of two naming tests and three drawing tests. One can see that the inputs in four out of the five tests are "new" to the system because they were not included in Table 2. In other words, the system has never had the chance of "knowing" them. The system would have made mistakes or made no response to the inputs if memory consolidation was a simple relocation of experience. Instead, it names those presented "pictures" without mistake, and counts the peaks of the representation inputs, and draws "pictures" correctly by tallying the symbol and drawing the peaks out accordingly. Although those four inputs seem "new" to the system, they are not new at all in meaning, or in the sense of common features and conceptual knowledge. The same common features (and conceptual knowledge) have already been carried by those input pairs in Table 2 and have already been abstracted and generalized into the learning center during the system's dreams.

**Conclusion**

The presented cognitive model of dreaming confirms computationally that learning and memory consolidation of episodic memory can be realized with randomly arriving signals that are stored segments of experience or episodic memory. This model offers a positive answer to Freud's inquiry (1900): "Is the dream capable of teaching us something new concerning our internal psychic processes and can its content correct opinions which we have held during the day?" It also demonstrates that dreaming is a cognitive process that deals with the "daily residual" (Freud, 1900), "unfinished business" (Hall & Van De Castle, 1966) and correspondences with waking thoughts (Domhoff, 1996; Foulkes, 1985). Although this system is only able to demonstrate the dream functions of learning and memory consolidation, there is no doubt that dreaming functions should cover many other cognitive aspects and psychological properties as many studies have concluded. The activation-synthesis theory states that, during dreaming, the brain "tries its best to make sense out of the nonsense." We may complete this statement by adding: when the brain does its best, it is able to make sense out of the nonsense of randomly arriving signals.

## References

Antrobus, J. (1987). Cortical hemisphere asymmetry and sleep mentation. *Physiological Review* 94:359-68.

Battaglia1, F.P., Sutherland, G.R., & McNaughton, B.L. (2004). Hippocampal sharp wave bursts coincide with neocortical "up-state" transitions. *Learning & Memory* 11:697-704.

Bloch, V., Hennevinm E., & Leconte, P. (1979). Relationship between paradoxical sleep and memory processes. In: Brazier MA (ed), *Brain mechanisms in memory and learning: from the single neuron to man* (pp 329-343), Raven Press: New York.

Botman, H.I., & Crovitz, H.F. (1989). Dream reports and autobiographical memory. Imagination, *Cognition and Personality* 9:213-214.

Bradley, L., Hollifield, M., & Foulkes, D. (1992). Reflection during REM dreaming. *Dreaming* 2:161-166.

Braun, A.R., Balkin, T.J., Wesensten, N.J., Carson, R.E., Varga, M., Baldwin, P., Selbie, S., Belenky, G., & Herscovitch, P. (1997). Regional cerebral blood flow throughout the sleep-wake cycle. *Brain* 120:1173-1197.

Christos, G. (2003). Memory and dreams: *The creative human mind*. Rutgers University Press: New Jersey.

Crick, F, & Mitchison, G. (1983). The Function of dream Sleep. *Nature* 304:111-115.

Domhoff, G.W. (1996). *Finding meaning in dreams: a quantitative app*roach. Plenum: New York.

Domhoff, G.W. (1993). The repetition of dreams and dream elements. In: Moffitt A, Kramer M, Hoffmann R (eds), *The functions of dreaming* pp (293-320), SUNY Press: Albany.

Fishbein, W. (1970). Interference with conversion of memory from short-term to long-term storage by partial sleep deprivation. *Communications in Behavioral Biology* 5:171-175.

Fosse, M.J., Fosse, R., Hobson, J.A., & Stickgold, R.J. (2003). Dreaming and episodic memory: a functional dissociation? *J. Cogn. Neurosci*. 15:1-9.

Foulkes, D. (1985). *Dreaming: a cognitive-psychological analysis*. Erlbaum: Hillsdale.

Foulkes, D. (1962). Dream reports from different states of sleep. *Journal of Abnormal and Social Psychology* 65:14-25.

Freud, S. (1900). *The interpretation of dreams*. Translated and edited by J. Strachey. Basic Books: New York.

Greenberg, R., & Pearlman, C. (1974). Cutting the REM nerve: An approach to the adaptive role of REM sleep. *Perspectives in Biology & Medicine* 17:513-521.

Hall, C. S. (1953). A cognitive theory of dream symbols. *Journal of General Psychology* 48:169-186.

Hall, C.S., & Van de Castle R.L. (1966). *The content analysis of dreams*. Appleton-Century-Crofts: New York

Hartmann, E. (1966). The psychophysiology of free will. In: Lowenstein E (ed), *Psychoanalysis: A General Psychology* (pp 521-536). International University Press: New York.

Hennevin, E., Hars, B., Maho, C., & Bloch, V. (1995). Processing of learned information in paradoxical sleep: relevance for memory. *Behavioral Brain Research* 69:125-135.

Hobson, J.A. (1988). *The Dreaming brain*. Basic Books: New York

Hobson, J.A., & McCarley, R.W. (1977). The brain as a dream-state generator: An activation-synthesis hypothesis of the dream process. *American Journal of Psychiatry* 134:1335-48.

Hobson, J.A., Pace-Schott, E., & Stickgold, R. (2000). Dreaming and the Brain: Towards a Cognitive Neuroscience of Conscious States. *Behavioral and Brain Sciences* 23:793-842.

Kant, I. (1800). *Logic*, translated by R. S. Hartman & W. Schwarz (1974). New York: Dover Publications.

LaBerge, S. (1985). *Lucid dreaming*. Ballantine: New York.

Levy, W.B. (1996). A sequence predicting CA3 is a flexible associator that learns and uses context to solve hippocampal-like tasks. *Hippocampus* 6:579-590.

Locke, J. (1690). *An essay on human understanding*. Republished in 1997. Peguin Books: London.

Louie, K., & Wilson, M.A. (2001). Temporally structured replay of awake hippocampal ensemble activity during rapid eye Mmovement sleep. *Neuron* 29:145-156.

Maquet, P., Peters, J.M., Aerts, J., Delfiore, G., Degueldre, C., Luxen, A., & Franck, G. (1996). Functional neuroanatomy of human rapid-eye-movement sleep and dreaming. *Nature* 383:163-66

McClelland, .JL., McNaughton, B.L., & O'Reilly, R.C. (1995). Why there are complementary learning systems in the hippocampus and neocortex: insights from the successes and failures of connectionist models of learning and memory. *Psychol Rev*. 102:419–57.

McCarley, R.W., & Hoffman, E. (1981). REM sleep dreams and the activation-synthesis hypothesis. *American Journal of Psychiatry* 138:904-912.

Muller, R.U., Kubie, J.L., & Ranck, J.B. (1987) .Spatial firing patterns of hippocampal complexspike cells in a fixed environment. *J. Neurosci*. 7:1935–50.

Myers, R.E., & Sperry, R.W. (1953). Interocular transfer of a visual forma discrimination habit in cats after section of the optic chaism and corpus callosum. *Anat. Rec*. 115:351-352.

Nielsen, T.A. (2000). A review of mentation in REM and NREM sleep: "Covert" REM sleep as a possible reconciliation of two models. *Behavioral and Brain Sciences* 23:851-866.

Nielsen, T.A., & Stenstrom, P. (2005). What are the memory sources of dreaming? *Nature* 437:1286-1289.

Nofzinger, E.A., Mintun, M.A., Wiseman, M.B., Kupfer, D.J., & Moore, R.Y. (1997). Forebrain activation in REM sleep: An FDG PET study. *Brain Research* 770:192-201.

O'Keefe, J., & Nadel, L. (1978). *The hippocampus as a cognitive map*. The Clarendon Press: Oxford.

Pavlides, C., & Winson, J. (1989). Influences of hippocampal place cell firing in the awake state on the activity of these cells during subsequent sleep episodes. *Journal of Neuroscience* 9:2907-2918.

Pearlman, C. (1971). Latent learning impaired by REM sleep deprivation. *Psychonomic Science* 25:135-136.

Poe, G.R., Nitz, D.A., McNaughton, B.L., & Barnes, C.A. (2000). Experience dependent phase-reversal of hippocampal neuron firing during REM sleep. *Brain Res*. 855:176–180.

Purcell, S., Mullington, J., Moffitt, A., Hoffman, R., & Pigea, R. (1986). Dream self-reflectiveness as a learned cognitive skill. *Sleep* 9:423-37.

Rechtschaffen, A. (1978). The single-mindedness and isolation of dreams *Sleep* 1:97-109.

Rechtschaffen, A., & Kales, A. (1968). *A manual of standardized terminology, techniques and scoring system for sleep stages of human subjects*. Public Health Service: Washington

Reinsel, R., Antrobus, J., & Wollman, M. (1992). Bizarreness in dreams and waking fantasy. In: Antrobus JS, Bertini M (eds), *The neuropsychology of sleep and dreaming* (pp 157-184), Lawrence Erlbaum Associates: Hillsdale.

Revonsuo, A., & Salmivalli, C. (1995.) A content analysis of bizarre elements in dreams. *Dreaming* 5:169-87.

Russell, B. (1913). *Theory of knowledge: The 1913 Manuscript*. Republished in 1984. Allen and Unwin: London

Schwartz, S. (2003). Are life episodes replayed during dreaming? *Trends in Cognitive Sciences* 7:325-327.

Scoville, W.B., & Milner, B. (1957). Loss of recent memory after bilateral hippocampal lesions. Journal of Neurology. *Neurosurgery and Psychiatry* 20:11–21.

Skaggs, W.E., & McNaughton, B.L. (1996). Replay of neuronal firing sequences in rat hippocampus during sleep following spatial experience. *Science* 271:1870–1873.

Smith, C. (1995). Sleep states and memory processes. *Behavioral Brain Research* 69:137-145.

Solms, M. (1997). *The neuropsychology of dreams*. Lawrence Erlbaum: Mahwah

Sperry, R. (1982). Some effects of disconnecting the cerebral hemispheres. *Science* 217:1223-1226.

Squire, L.R. (1992.) Memory and the hippocampus: A synthesis from findings with rats, monkeys, and humans. *Psychological Review* 99:195-231.

Squire, L.R., & Alvarez, P. (1995). Retrograde amnesia and memory consolidation: a neurobiological perspective. *Current Opinion in Neurobiology* 5:169-177.

Staba, R.J., Wilson, C.L., Fried, I., & Engel, J.J. (2002). Single neuron burst firing in the human hippocampus during sleep. *Hippocampus* 12:724-734.

Stickgold, R., Hobson, J.A., Fosse, R., & Fosse, M. (2001). Sleep, learning, and dreams: off-line memory reprocessing. *Science* 294:1052–1057.

Tulving, E. (1972). Episodic and semantic memory. In: Tulving E, Donaldson W (eds), *Organization of memory* (pp 381-403), Academic Press: New York

Williams, J., Merritt, J., Rittenhouse, C., & Hobson, J.A. (1992). Bizarreness in dreams and fantasies: Implications for the activation-synthesis hypothesis. *Consciousness and Cognition* 1:172-85.

Wilson, M.A., McNaughton, B.L. (1994). Reactivation of hippocampal ensemble memories during sleep. *Science* 265:676-679.

Winson, J. (1985). *Brain and psyche: The Biology of the Unconscious*. Doubleday/Anchor Press: New York.

Wolf, F.A. (1994). *The dreaming Universe*. Simon & Schuster: New York.

Zhang, Q. (2005). An artificial intelligent counter. *Cognitive Systems Research* 6:320-332.

## Figures and captions

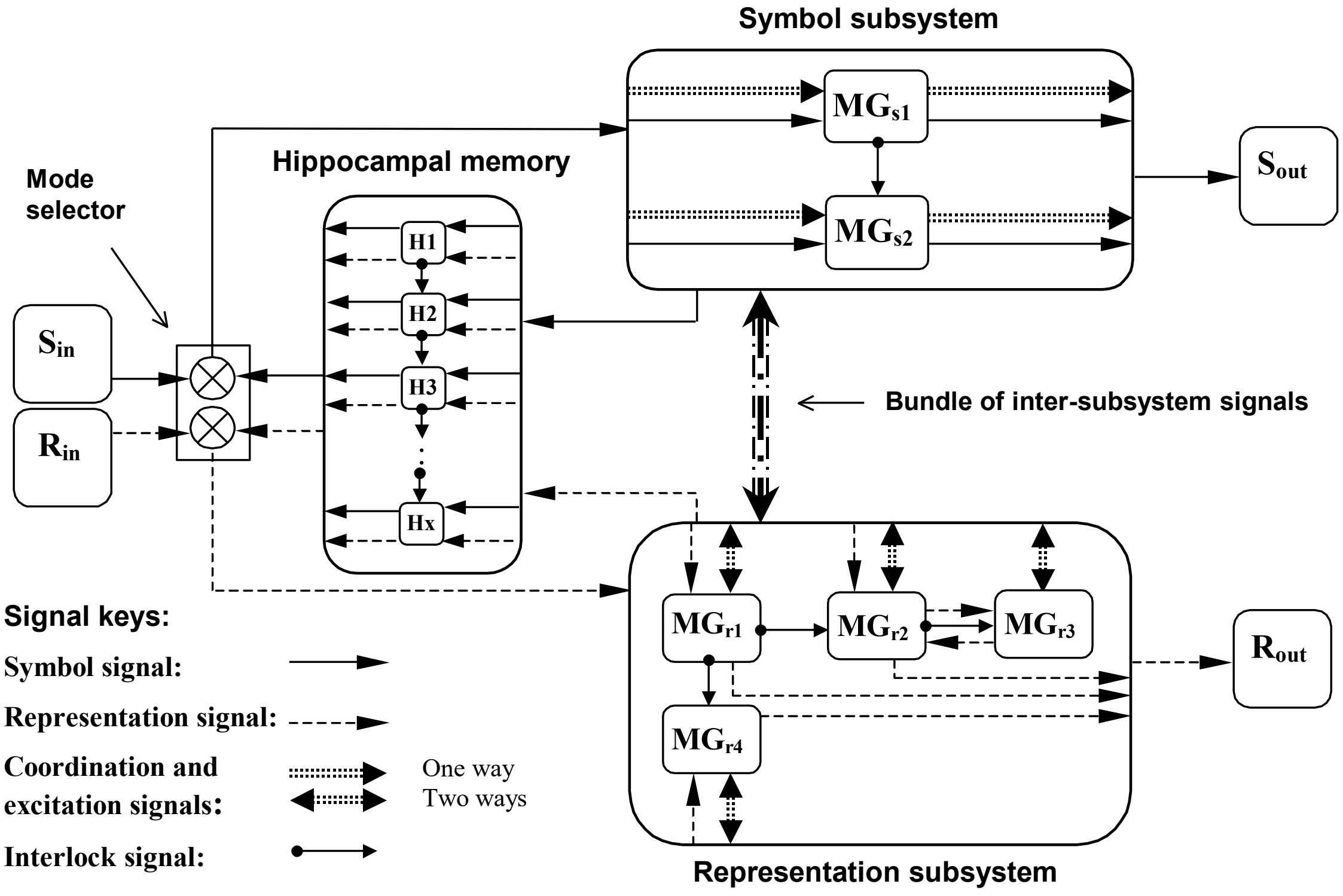

Figure 1. The cognitive system of a computational dreamer. This system has two cognitive sections. One is the learning center (or the AI counter) that learns knowledge from external experiences; the other is the hippocampal memory that stores unlearned experiences. The learning center is made up of two subsystems, the symbol and representation subsystems. Each of the subsystems has several memory triangles (e.g., $MG_{r1}$) that are interlocked to and associated with each other by interlocking signals, excitation signals and coordination signals. The pathways and kinds of signals are indicated under the "Signal keys". The hippocampal memory, consisting of a number of interlocked memory cells, is designated to store those external experiences from which the learning center cannot learn. The $S_{in}$ and $R_{in}$ are input ports of a symbol and representation, respectively; while $S_{out}$ and $R_{out}$ are output ports. The mode selector can set the system to either waking mode (for external signals) or dreaming mode (for random internal signals).

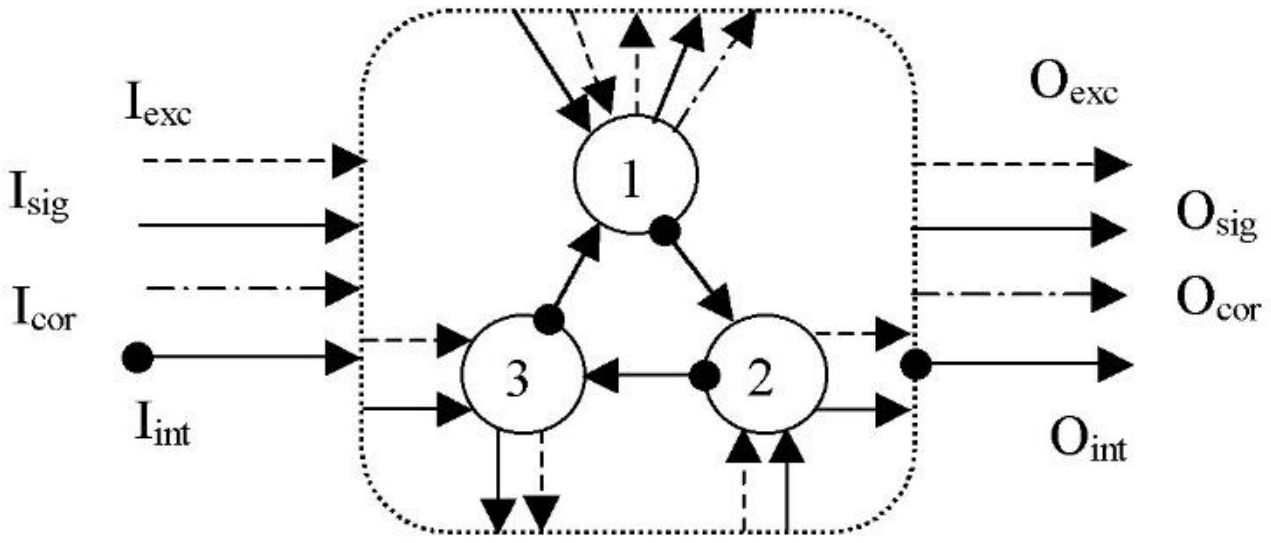

Figure 2.  A memory triangle that is made up of three single memories (after Zhang, 2005).  $I_{exc}$ is excitation input, $O_{exc}$ is excitation output, $I_{sig}$ is signal input, $O_{sig}$ is signal output, $I_{cor}$ is coordination input, $O_{cor}$ is coordination output, $I_{int}$ is interlock input, and $O_{int}$ is interlock output.

## Tables

Table 1. Sequenced Input pairs

| Order | Symbol | Representation |
|---|---|---|
| | For common feature of “zero” | |
| 1st | Z | 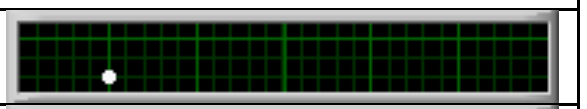 |
| 2nd | Z | 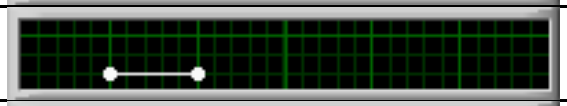 |
| 3rd | Z | 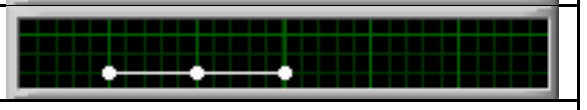 |
| | For common feature of “one” | |
| 4th | I | 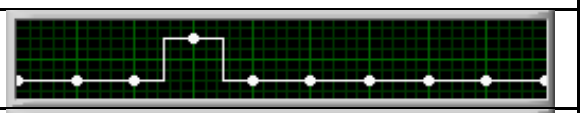 |
| 5th | I | 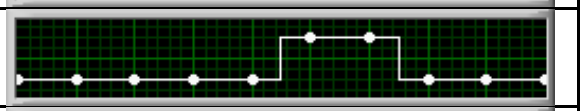 |
| 6th | I | 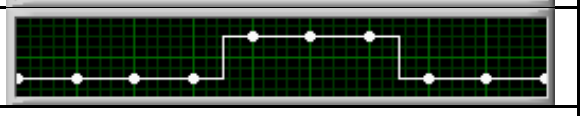 |
| | For common feature of tallying | |
| 7th | II | 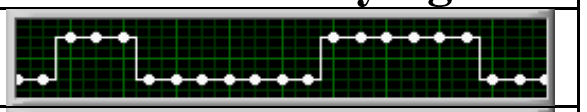 |
| 8th | III | 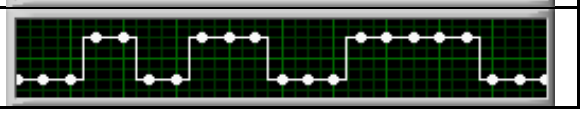 |

Table 2. Non-sequenced input pairs

| Order | Symbol | Representation |
|---|---|---|
| 1st | III | 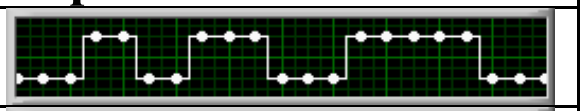 |
| 2nd | I | 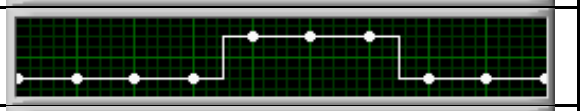 |
| 3rd | I | 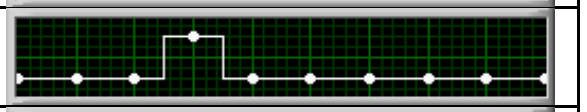 |
| 4th | Z | 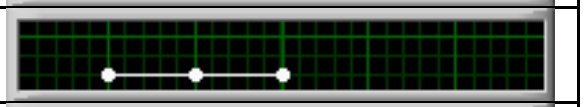 |
| 5th | Z | 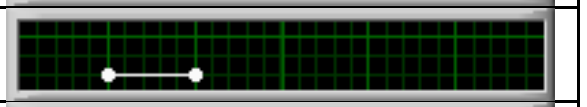 |
| 6th | Z | 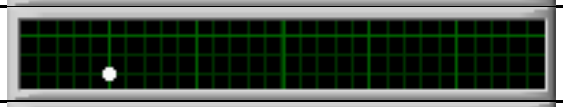 |
| 7th | II | 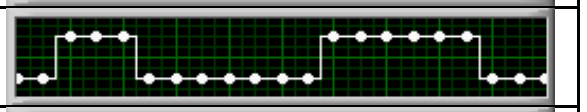 |
| 8th | I | 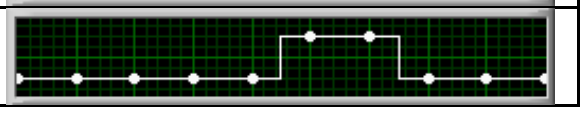 |

Table 3. Recorded streams of firings from the hippocampal memory

| Mode | Recorded stream of representation firings | Result |
|---|---|---|
| Waking: Serial recall | 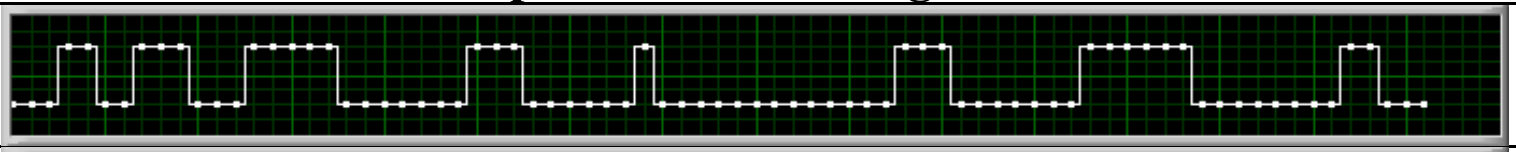 | Experience is recalled |
| Dreaming: One dream | 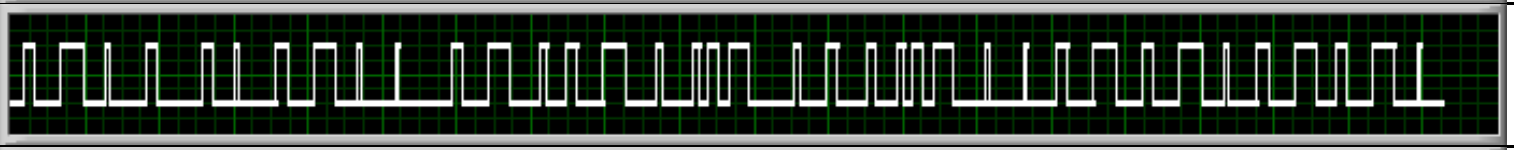 | Can perform tally |
| Dreaming: One dream | 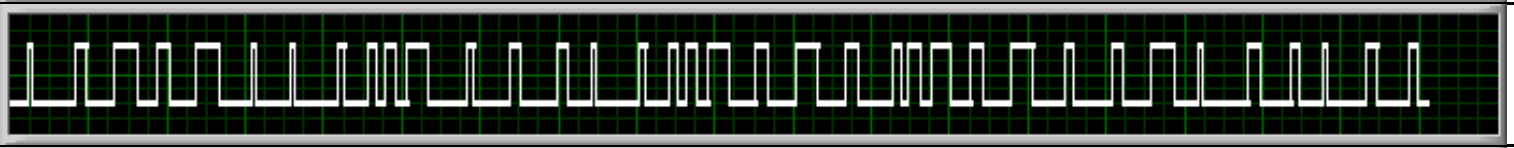 | Can perform tally |
| Dreaming: Four dreams | 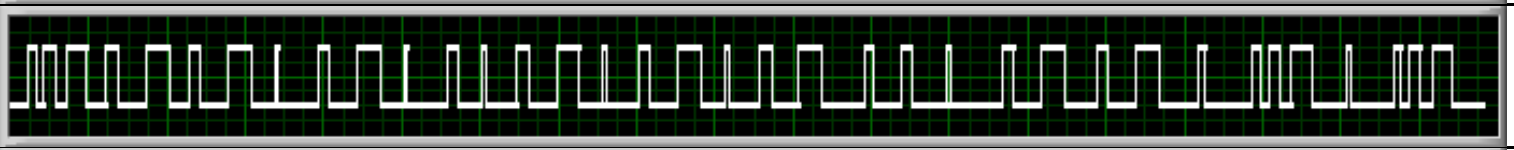 | Cannot fully perform tally |
| | 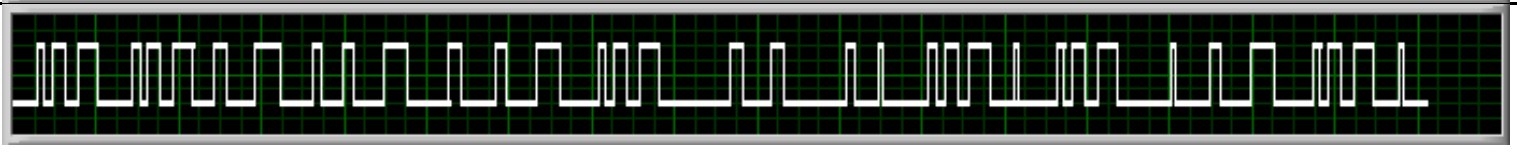 | Cannot fully perform tally |
| | 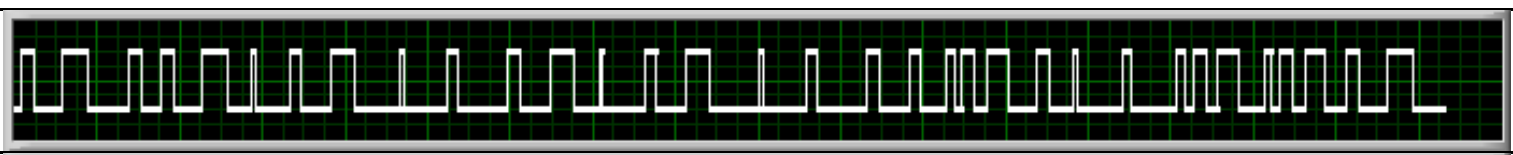 | Cannot fully perform tally |
| | 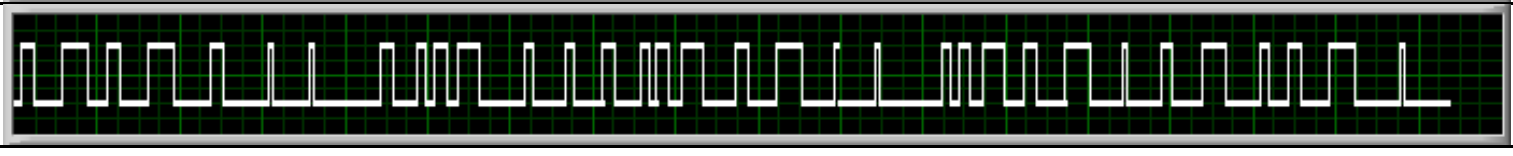 | Can perform tally |

Table 4. The naming and picture drawing tests

| Input | | Output |
|---|---|---|
| R | 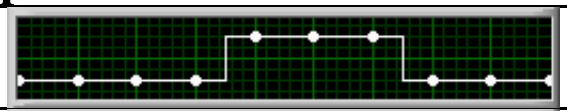 | I |
| R | 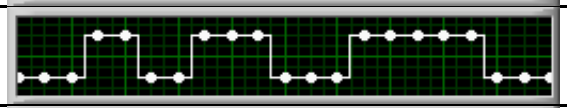 | III |
| R | 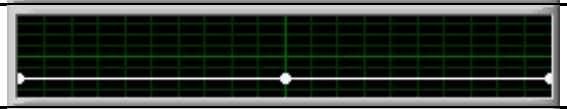 | Z |
| R | 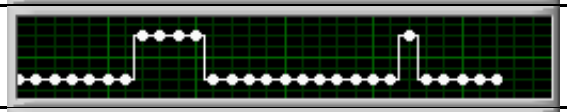 | II |
| R | 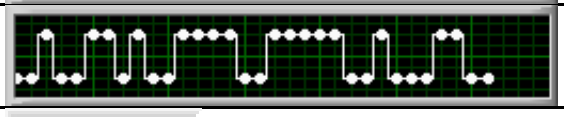 | IIIIIII |
| S | Z | 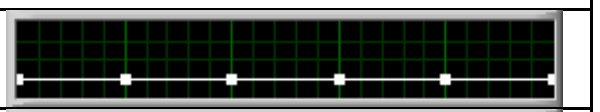 |
| S | IIIII | 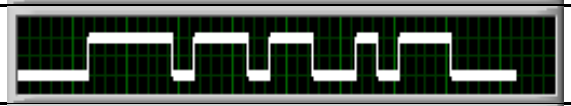 |
| S | IIIIIII | 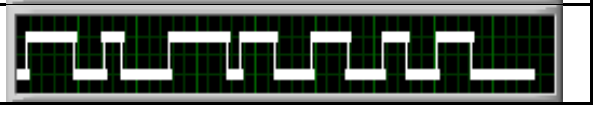 |

R: representation input; S: symbol input